# Towards an Effective Intrusion Response Engine Combined with Intrusion Detection in Ad Hoc Networks[†]

Aikaterini Mitrokotsa *, Nikos Komninos** and Christos Douligeris*
* Department of Informatics, University of Piraeus, Piraeus, Greece
** Athens Information Technology, Attiki, Greece
{mitrokat, cdoulig}@unipi.gr, nkom@ait.edu.gr

*Abstract* — **In this paper, we present an effective intrusion response engine combined with intrusion detection in ad hoc networks. The intrusion response engine is composed of a secure communication module, a local and a global response module. Its function is based on an innovative tree-based key agreement protocol while the intrusion detection engine is based on a class of neural networks called eSOM. The proposed intrusion response model and the tree-based protocol, it is based on, are analyzed concerning key secrecy while the intrusion detection engine is evaluated for MANET under different traffic conditions and mobility patterns. The results show a high detection rate for packet dropping attacks.**

*Keywords* – Ad Hoc Networks, Intrusion Response, Intrusion Detection, Key Agreement Protocol.

## I. INTRODUCTION

There are many "flavours" of ad hoc networks. In this paper, we refer to an open ad hoc network composed of a set of nodes that are characterized by short membership duration since there is a large number of joining and leaving events. In order to achieve security in ad hoc networks, complementary to intrusion prevention techniques there is a need for reactive mechanisms, such as intrusion detection and response.

Intrusion Detection is a mature arsenal for the defense of wired networks, but it is still in its infancy in the area of ad hoc networks. Nevertheless, there are some proposed IDS systems for ad hoc networks. Zhang and Lee [1] proposed the first (high-level) specific for ad hoc networks IDS approach. They proposed a distributed and cooperative anomaly-based IDS, which provides an efficient guide for the design of IDS in wireless ad hoc networks. Huang and Lee [2] extended their previous work in a cluster-based IDS, in order to combat the resource constraints that MANET face. Liu et al. [3] proposed a completely distributed anomaly detection approach. They investigated the use of the MAC layer in order to profile normal behavior of mobile nodes and then applied cross-feature analysis on feature vectors constructed from the training data.

Furthermore, effective intrusion detection should be combined with an efficient and secure intrusion response. Intrusion response in order to be efficient and secure should be based on security mechanisms such as group key management and agreement.

However, most of the proposed key management protocols ([4], [5], [6]) for wired networks can not be applied in an infrastructure-less or resource sensitive environment such as ad hoc networks. In the Octopus protocol [4], four nodes compose a $2^2$ −cube and the rest members of the network are "tentacles" that are connected to one of the central nodes. In a dynamic ad hoc network, it is not easy to maintain such a topology. The Tree-Group Diffie-Hellman (TGDH) proposed by [5], which is based on a binary tree structure and improves the performances of the IKA1/2 [6], is based on modular exponentiation which is the most expensive operation thus, it might require O(n) exponentiations in order to compute the group session key.

Hwang and Chang [7] have proposed a key agreement protocol that is based on a shared group password, the exclusive-OR (XOR) operation and a binary tree structure. Although this approach is really efficient for ad hoc networks it is susceptible to password guessing and replay attacks. Lo [8] et. al. have improved the previous approach [7] by adding mutual authentication and adding a procedure for periodic session key updates. However, their approach is still susceptible to dictionary and brute force attacks since it is based on passwords.

In this paper, we propose an Intrusion Response engine and an Intrusion Detection engine based on a class of neural networks called emergent Self-Organizing Maps (eSOM), in order to ensure the direct response to possible attacks and the effective exploitation of the information visualization that eSOM provide, for every node. The proposed Intrusion Response engine is based on an authenticated key agreement protocol, which has some unique characteristics including the use of a shared master key and the structure of a rooted tree, which depends on the distance between the ad hoc nodes and consequently a unique protocol flow. Using this protocol we are able to generate Local Keys (LK) and a Global Key (GK) in order to secure the communication of the proposed Intrusion Response engine.

---

[†] This work was partially supported by the Greek Research Technology Secretariat under a PENED grant and the project CASCADAS (Componentware for Autonomic Situation-aware Communications, and Dynamically Adaptable Services), which is funded by IST FET Program of the European Commission (contract FP6-027807).

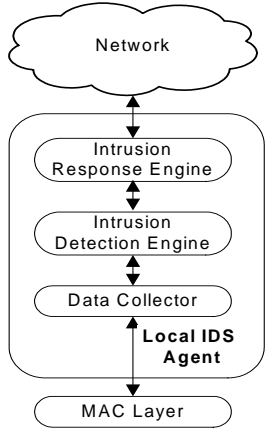

Fig. 1. Intrusion Detection Architecture

## II. INTRUSION DETECTION MODEL

The IDS architecture we adopt is composed of multiple local IDS agents, as illustrated in Fig. 1, that are responsible for detecting possible intrusions locally. The collection of all the independent IDS agents forms the IDS system for the ad hoc network. Each local IDS agent is composed of the following components:

*Data collector*: is responsible for selecting local audit data and activity logs.

*Intrusion Detection engine:* is responsible for detecting local anomalies using local audit data. The local anomaly detection is performed according to the following procedure:
- Select labeled audit data and perform the appropriate transformations.
- Compute the classifier using training data and the eSOM algorithm.
- Apply the classifier to test local audit data in order to classify it as normal or abnormal.

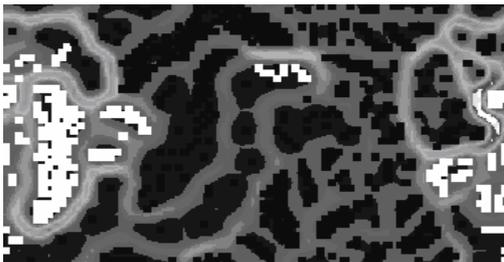

Fig. 2. Emergent SOM U-Matrix of a node of an ad hoc network

An example of an eSOM map is illustrated in Fig. 2. We have to note here, that the integrity of the generated eSOM map is assured with integrity mechanisms such as a MAC (Message Authentication Code) or a hash function.

*Intrusion Response Engine:* If the *Intrusion Detection Engine* detects an intrusion then the *Intrusion Response Engine* is activated. Each node of the ad hoc network is able to participate in the *Intrusion Response Engine*. The *Intrusion Response Engine* is responsible for sending a local and a global alarm in order to notify the nodes of the ad hoc network about the incident of an intrusion. The proposed Intrusion Response Engine is composed of three main modules: the *Communication Module*, the *Local Response Module* and the *Global Response Module*. The *Local Response Module* is activated every time an intrusion is detected by the Intrusion Detection Engine while the *Global Response Module* is activated only in serious cases of attacks i.e the eSOM map of a node (Fig. 2) is covered in its biggest part (over two thirds (2/3)) with signs of attack (light color).

## III. PROPOSED INTRUSION DETECTION ENGINE

The Intrusion detection engine is based on emergent Self Organizing maps [9] a special class of neural networks. We have used the distance based (U-Matrix) method in order to visualize the structures generated by eSOM. The U-Matrix is a display of the U-heights on top of the grid positions of the neurons on the map. The input data set is displayed and depicted at a 3D landscape. The height will have a large value in areas of the map where one finds a few data points and small value in areas that represent clusters, creating hills and valleys correspondingly.

In our examples, we trained eSOM with logs of network traffic selected from a simulated MANET and used eSOM U-matrices [9] to perform intrusion detection. A vector represents each log of network traffic with some fixed attributes. Each vector has a unique spatial position in the U-Matrix while the distance between two points is the dissimilarity of two network traffic logs. The U-Matrix of the trained dataset is divided into valleys that represent clusters of normal or attack data and hills that represent borders between clusters. Depending on the position of the best match of an input data point that characterizes a connection, this point may belong to a valley (cluster (normal or attack behavior)) or this data point may not be classified if its best match belongs to a hill (boundary). The map that is created after the training of the eSOM, will represent the network traffic. Thus, an input data point may be classified depending on the position of its best match. We exploit the advantages of key agreement protocols in order to verify the authenticity and integrity of the maps.

## IV. PROPOSED INTRUSION RESPONSE ENGINE

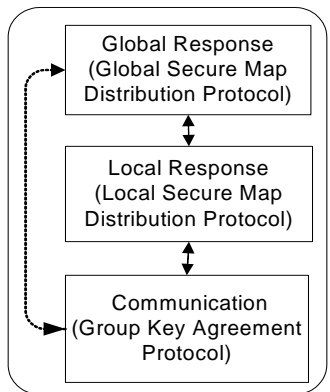

Fig. 3. Intrusion Response Engine

The Intrusion Response engine and its components are illustrated in Fig. 3. The *Communication Module* is responsible for the agreement of the local keys (LK) and a global key (GK) that will be used in the local and global response modules, respectively.

The *Local Response Module* is responsible for the secure distribution of the global eSOM map, generated by a node using its eSOM map and the local eSOM maps of all its one-hop away neighbors. In order to verify the integrity and authenticity of the global eSOM map we apply a *Local Map Distribution Protocol*. The generated global eSOM map is used for the visualization of the security status of the local ad hoc network composed of one-hop neighbors of a node. This map also helps nodes to select the most appropriate and secure neighboring node for message forwarding. The *Global Response Module* is responsible for the notification of all neighbors in the communication range of the attacked node.

### A. Communication module

We propose an authenticated group key agreement protocol, which is based on the group key agreement protocols proposed by Lo et. al. [8] and Huang and Chang [7]. Our protocol adopts the exclusive-OR (XOR) operation proposed by [7], [8] and not modular or exponential calculations that have extremely high computational cost. Furthermore, it employs some unique characteristics that make it even more secure. It uses a shared secret master key ($K_M$) and it is based on the structure of a rooted tree which depends on the distance between the ad hoc nodes. This results in a unique protocol flow.

In the following paragraphs we give a short description of the way the Group Key Agreement (GKA) protocol functions. Table I provides the description for every notation used in the description of the GKA protocol and the local and global response modules. Our authenticated key agreement protocol is not based on passwords as in [7] and [8], since such protocols are susceptible to dictionary and brute force attacks. In our scenario, there are *n* members sharing a secret master key $K_M$. $K_M$ is shared among all nodes in the ad hoc network and it is used for the initial communication. Although $K_M$ can be used as a first step in order to setup a secure communication it is neither efficient nor sufficient to be used for a secure session communication. In addition, our proposed key agreement protocol is not based in a complete binary tree as in [8] but in a simple rooted tree.

TABLE I
GKA ALGORITHM NOTATIONS DESCRIPTION

| Notation | Description |
|---|---|
| $M_i$ | Member i |
| $M_d$ | The member in a descendant node |
| $M_a$ | The member in an ascendant node |
| $M_{at}$ | The member of the ad hoc network that is the victim of an attack |
| $M_{C_j}$ | The member of the ad hoc network that is the jth child of its parent node in the tree structure |
| $M_L$ | The member of the ad hoc network that leaves the ad hoc network (leave protocol) |
| $ID_i$ | Member i's identity |
| $K_M$ | Master Key |
| H( ) | One – way hash function |
| $S_i$ | Member i's contributory key |
| z | Subkey generated by $M_1,...,M_{n-1}$ |
| $K_i'$, $K_i$ | $K_i'$ is the intermediate key and $K_i$ is the session key hold by $M_i$. |
| $nonce_i$ | Member i generated random number |
| $T_i$ | The tree path of member node $M_i$ along its parent node to root node (ex. in Figure 2 the key path $T_5$ of node $M_5$ is $M_5 \rightarrow M_2 \rightarrow M_1$) |
| $\oplus$ | XOR operation |
| $\|$ | Concatenation |
| $map_i$ | The eSOM map of member node i |
| LK | Local secret Key |
| GK | Global secret Key |
| Br. | The corresponding message is broadcasted. |

We assume that there are n members $M_1, M_2, …, M_n$, in an ad hoc network that want to have a secure communication. Initially, each member of this group has a unique identity number $ID_i$, for i= 1, 2, ..., n. These members cooperate based on a rooted tree. In this rooted tree structure, every node is either a leaf or a parent of one or more children nodes. The nodes are denoted with each member's unique number. In this group, we assign member $M_{Ch}$ to be a "Checker". The checker is a group member that is randomly selected between the one-hop neighbors of the root node. The checker does not participate in the tree structure, but has an additional role to confirm the session key correctness. In case that the randomly selected $M_{Ch}$ leaves the group of nodes due to mobility reasons, the root node is able to detect this movement and select another one-hop neighbor as a "checker".

The rooted tree is constructed based on the distance between nodes and their unique numbers. If we assume that level 0 is the root of the tree and it is situated in node $M_1$ with ID number $ID_1$, then in level one will be situated its one-hop neighbors; in level two its two-hop away neighboring nodes until the last level where the most distant nodes of node $M_1$ that are in its transmission range, will be situated. Fig. 4 illustrates an example of a key tree with 17 members. In the key tree, its root is located at level l=0 and its height is h, where h=4.

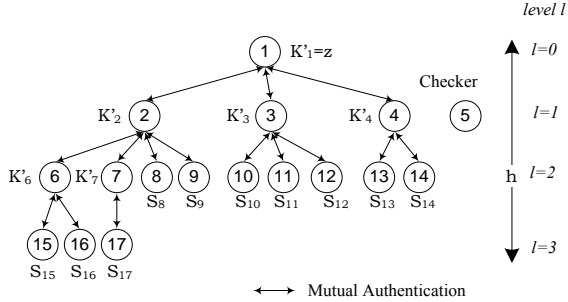

Fig. 4. Key structure illustrates membership - Subkey z generation by $M_1$ to $M_{17}$.

Our goal is all nodes in the ad hoc network to agree upon a session key K:
$$K = S_1 \oplus S_2 \oplus .... \oplus S_n, \quad (2)$$
where $S_i$ is contributed by $M_i$ and it is randomly selected. The protocol is divided into two phases: the *key initiation phase* and *the session key generation phase*. In the *key initiation phase*, $M_1, M_2, ..., M_n$ (we except the checker node $M_{Ch}$) cooperate to secretly construct a subkey z: $z = S_1 \oplus S_2 \oplus ... \oplus S_n$. (3)

In the session key generation phase, each $M_i$ ($i= 1, 2, ..., n$ and $i \neq ch$) engages in a separate exchange with $M_{Ch}$. After this exchange all members have sufficient information to compute the session key $K$. Member node $M_{Ch}$, also verifies that the other members generated the same session key $K$. We introduce our method in detail in the following subsections.

*1) Key Initiation Phase*

Initially, all member nodes are considered to be malicious unless proven otherwise. A node that starts the normal communication process (key initiation phase) is considered to be valid. The proposed key initiation phase follows a unique protocol flow based on the structure of a rooted tree depending on the distance of the ad hoc nodes. Each member $M_i$ chooses a random quantity $S_i$. For $i \neq ch$, all member nodes perform three steps to achieve mutual authentication. The protocol flow of the *Key Initiation Phase* is the following:

For ($i \neq ch$): Steps 1 to 3 are used for mutual authentication.

**Step 1**
$$M_d \xrightarrow{ID_d, ID_a, E_{K_M}(ID_d \| ID_a \| nonce_d)} M_a$$

**Step 1**
$$M_d \xrightarrow{ID_d, ID_a, E_{K_M}(ID_d \| ID_a \| nonce_d)} M_a$$

**Step 2**
$$M_d \xleftarrow{ID_a, ID_d, E_{K_M}(ID_a \| ID_d \| nonce_d + 1 \| nonce_a)} M_a$$

**Step 3**
$$M_d \xrightarrow{ID_a, ID_d, E_{K_M}(ID_a \| ID_d \| nonce_a + 1 \| K_i')} M_a$$

- If $M_i$ is a leaf node then $K_i' = S_i$. (4)
- If $M_i$ is a parent node and has one or more children represented as $C_j$ ($j= 1, ..., l$), then:
$$K_i' = S_i \oplus K'_{C_1} \oplus ... \oplus K'_{C_l}. \quad (5)$$
- If $M_i$ is the root node ($i=1$) then $M_1$ computes the subkey z
$$\begin{aligned}z &= S_1 \oplus K'_{C_1} \oplus K'_{C_2} \oplus ... \oplus K'_{C_j} \\ &= S_1 \oplus S_2 \oplus ... \oplus S_n\end{aligned} \quad (6)$$

An example of the subkey $z$ generation mechanism for nodes $M_i$ for i=1,2,...,17 and $i \neq 5$ is illustrated in Fig. 4. At the end of this phase the local keys LK of the one-hop neighbors of the root node are calculated and are used in the Local Response module: $LK = LK_j = z \oplus S_j$ (7) where j are all the one-hop neighbors of the root node and consequently belong in the first level of the tree.

*2) Session key agreement phase*

The subkey $z(= S_1 \oplus S_2 \oplus ... \oplus S_n)$ generated by $M_1$ at the end of the key initiation phase is used in the session key agreement phase. The protocol flow of the *Key Agreement Phase* is the following:

**Step 1**
$$M_1 \xrightarrow{\textbf{Br.}\ ID_1, E_{K_M}(ID_1 \| z \| nonce_1)} M_i,$$
$$\text{for } i=2, 3, ..., n$$

**Step 2**
$$M_{Ch} \xrightarrow{\textbf{Br.}\ ID_{Ch}, E_{K_M}(ID_{Ch} \| S_{Ch} \| nonce_i + 1 \| nonce_{Ch})} M_i$$
$$\text{for } i= 1, 2, ..., n \text{ and } i \neq ch$$

**Step 3**
$$M_{Ch} \xleftarrow{ID_i, ID_{Ch}, H(ID_{Ch} \| nonce_{Ch} + 1 \| K_i)} M_i$$

**Step 4**
$M_{Ch}$ checks session key K.

The session key K agreed in this phase is the Global Key (GK) that will be used in the Global Response module: $GK = K = z \oplus S_{Ch}$. (8)

*3) Membership events*

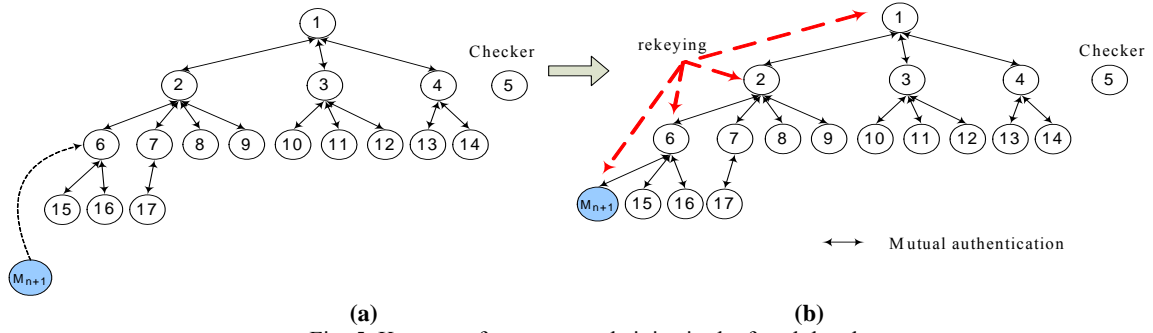

**(a)** **(b)**
Fig. 5. Key tree after a new node joins in the fourth level.

In a dynamic ad hoc network member nodes may join or leave the group, after the session key is generated. The new members are not authorized to know the session key agreed before they join the ad hoc network. Thus, the member nodes of the ad hoc network should change the agreed session key and the shared secret master key also. In the same manner, in case a node leaves the ad hoc network the session key should be changed in order to verify the secure communication of the remaining nodes. Thus, we need a Member Joining and a Member Leaving Protocol. This section describes the Member Joining Protocol. The Member Leaving Protocol is similar to the one described in [7] although it follows our new protocol flow and the structure of a simple rooted tree. Due to limited space it is not included in this paper but it is available in the extended version of the paper [10].

*Member Joining Protocol*

The proposed member joining protocol follows the protocol flow implied by the structure of the rooted tree. Let us assume that the group has $n$ members: $M_1, M_2, …, M_n$ and a new member $M_{n+1}$ wants to join a group of nodes. The new member sends a joining request message which includes its ID ($ID_{n+1}$). The new member will be allowed to join the group when the members in the group receive this message and accept it. Furthermore, the members of the group must change the master key from $K_M$ to $K'_M$ and reconstruct the session key.

The procedure that will be followed depends on how close to the root node $M_1$ the new member node $M_{n+1}$ will be. Thus, if the new member node $M_{n+1}$ is a one-hop neighbor of the root node $M_1$ then it will be situated in level one, if it is a two-hop neighbor of the root node $M_1$, and consequently one-hop neighbor node of a parent node in level one, it will be situated in level two. If we suppose that the new member node $M_{n+1}$ is $l$-hops away from root node $M_1$, then its position in the rooted tree will be in the $l^{th}$ level and would be a child of the node of which it is a one-hop neighbor. Then the member nodes in the key path $M_{n+1} \rightarrow M_{l-1} \rightarrow M_{l-2} \rightarrow … \rightarrow M_1$ where $M_{l-1}(M_{l-2})$ represents the member node that is situated in

$l$-$1^{st}$ ($l$-$2^{nd}$) level and it is parent node one (two)-hop neighbor of the new member node (respectively), perform the key initiation phase.

An example of a new member node $M_{n+1}$ joining a group of n members $M_1, M_2, …,M_n$ is illustrated in Fig. 5 where the new member will become a leaf node. The new member node $M_{n+1}$ is a three-hop neighbor of a root node $M_1$, a two-hop neighbor of member node $M_2$ and a one-hop neighbor of member node $M_6$. Each of the members $M_i$, except the root $M_1$ on the key-path $T_{n+1}$, $M_{n+1} \rightarrow M_6 \rightarrow M_2 \rightarrow M_1$ performs the key initialization phase. Particularly, if $M_d$ is a descendant node and $M_a$ is an ancestor node in the key path then the following protocol flow of the *Tree Path Key Initiation phase* is performed:

**Step a**
$$M_d \xrightarrow{ID_d, ID_a, E_{K'_M}(ID_d \| ID_a \| nonce''_d)} M_a$$

**Step b**
$$M_d \xleftarrow{ID_a, ID_d, E_{K'_M}(ID_a \| ID_d \| nonce''_d+1 \| nonce''_a)} M_a$$

**Step c**
$$M_d \xrightarrow{ID_a, ID_d, E_{K'_M}(ID_a \| ID_d \| nonce''_a+1 \| K''_i)} M_a$$

- If $M_i$ is a leaf node, then $K''_i = S''_i$ . (9)
- If $M_i$ is a parent node and has one or more children represented as $C_j$ ($j= 1,…,l$), then
  $$K''_i = S''_i \oplus K''_{C_1} \oplus … \oplus K''_{C_l} .\qquad(10)$$
- If $M_i$'s child $M_{C_j}$ is not in the key path, then
  $$K''_{C_j} = K'_{C_j} ,\qquad(11)$$
- If $M_i$ is the root node ($i=1$), then $M_1$ computes the subkey $z'$ using the $K''_i$ of the members in the key path and the $K_i$ of the other members of the tree and computes the new sub key:
  $$z' = S_1 \oplus K'_{C_1} \oplus K'_{C_2} \oplus … \oplus K'_{C_j}\qquad(12)$$

Finally, $M_1$ performs the algorithm of the key agreement phase (section IV.A.2) to reconstruct and verify a new session key.

*4) Periodic Session Key Update*

If no member joining or leaving events take place for a long period of time, the session key K,

must be changed in order to prevent it from possible exposure and to verify its security strength. The protocol flow of the *Periodical Global Session Key Update* follows the one described in [8]. We note here that the new session key $K_{new}$ is the new Global Key ($GK_{new}$) that will be used in the Global Response Module: $GK_{new} = K_{new} = K_{old} \oplus S''_{Ch}$ (13)

Except of the periodical update of the Global Key the Local Keys used by the Local Response module should also be updated after a long period of time. The protocol flow of the *Periodical Local Keys Update* is the following:

**Step 1**

$$M_j \xrightarrow{ID_j, E_{LK_{old}}(ID_j \| S''_j \| nonce_j)} M_1$$

where j is a node in level 1 of the tree

**Step 2**

$M_j$, computes the new Local Keys,

$$LK_{new} = LK'_j = LK_{old} \oplus S''_j \quad (14)$$

**Step 3**

$$M_j \xrightarrow{ID_j, H(ID_j \| nonce_j +1 \| LK'_j)} M_1$$

**Step 4**

$M_1$ checks the new Local Keys $LK_{new} = LK'_j$.

### B. Local Response Module

The *Local Response Module* is responsible for generating the eSOM map of the one-hop node connectivity network. More precisely, in a local ad hoc network each node creates its local eSOM map through the *Intrusion Detection Engine*. Each node transfers its local eSOM map, through the *Local Map Distribution Protocol*, to all its one-hop neighbors. The way the *Local Map Distribution Protocol* functions is illustrated in Fig. 6.

Suppose that we have a group of ad hoc nodes $M_1,...,M_n$ and $M_i$, for i=2, 3, ..., n, that are one-hop away neighbors of node $M_1$, then the protocol flow of the *Local Map Distribution Protocol* is the following:

**Step 1**

$$M_1 \xrightarrow{\textbf{Br.} \; ID_1, map_1, H_{LK}(ID_1 \| map_1 \| nonce_1)} M_i$$
$$for \; i=2,3,...,n$$

**Step 2**

$$M_1 \xleftarrow{ID_i, map_i, H_{LK}(ID_i \| nonce_1 +1 \| map_i)} M_i$$
$$for \; i=2, 3, ..., n$$

**Step 3**

$M_1$ creates the local map of one-hop neighbors

**Step 4**

$$M_1 \xrightarrow{\textbf{Br.} \; ID_1, GlobMap, H_{LK}(ID_1 \| GlobMap \| nonce_1)} M_i$$
$$for \; i=2, 3, ..., n$$

In Fig.6, nodes C, B, D and G are one-hop neighbors of node A. Nodes B, C, D and G create their own eSOM U-Matrix and use the local map distribution protocol in order to transfer securely their map to node A. Node A selects the local authenticated eSOM U-Matrices from its neighbors, creates its own map and then creates the global map of its local network. By observing the global map of its local network, node A is able to have a view of the security status of its neighboring nodes. Based on this information node A selects the appropriate node in order to forward its messages. By observing the local maps of all its neighboring nodes and by considering as secure the nodes that are not victims of attacks, node A selects an appropriate node for message forwarding.

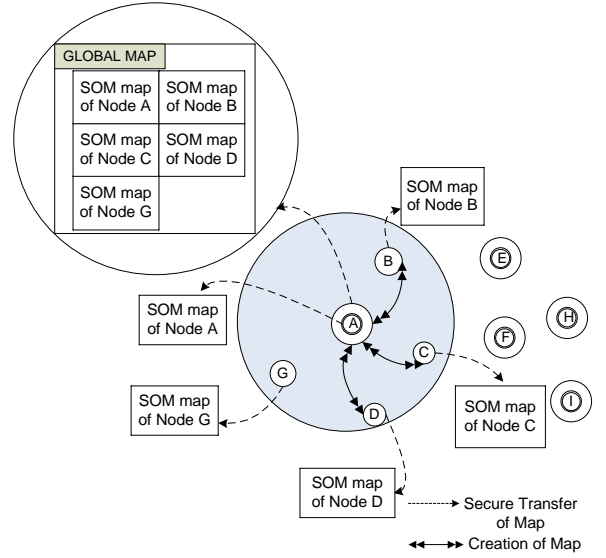

Fig. 6. Functioning of Local Response Module

### C. Global Response Module

The *Global Response Module* is responsible for notifying all the neighbors of the attacked node $M_{at}$, not only the ones that are one-hop away but all nodes in its transmission range ($M_i$, for i=1, 2, ..., n and i≠at), that there is a possible intrusion in this node. The notification of all nodes in the transmission range of the attacked node is performed using the *Global Response - Map Distribution Protocol*.

Suppose that we have a group of ad hoc nodes $M_1, ..., M_r$ and $M_i$, for i =2, 3, ..., r, that are all nodes in the transmission range of member node $M_1$ and $M_{at}$ is a member node that is a possible victim of the attack then the protocol flow of the *Global Response-Map Distribution Protocol* is the following:

**Step 1**

$$M_{at} \xrightarrow{\textbf{Br.} \; ID_{at}, H_{GK}(ID_{at} \| map_{at} \| nonce_{at})} M_i$$
$$for \; i=1, 2, ..., n \; and \; i \neq at$$

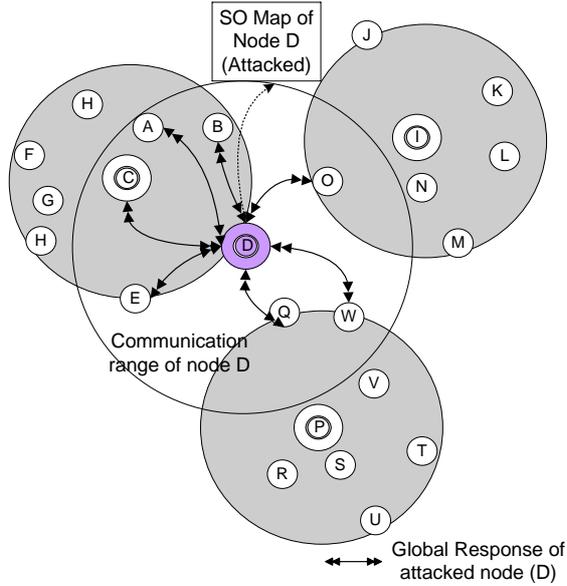

Figure 7. Functioning of Global Response Module

**Step 2**

$M_i$, for $i=1, 2, ..., n$ and $i \neq at$ removes $M_{at}$ from its routing tables.

The keyed hash function uses the global key GK generated by the communication module:

$$GK = K = z \oplus S_{Ch}. \tag{15}$$

Fig. 7 depicts an example of how the *Global Response Module* functions. In this figure three local ad hoc networks are illustrated. One with one-hop neighbors for node C (including nodes A, B, D, E, F, G, H), one with one-hop neighbors for node I (including nodes J, K, L, M, N, O), and one with one-hop neighbors for node P (including nodes Q, R, S, T, U, V, W). We suppose that the *Intrusion Detection Engine* has detected node D as the victim of an attack. Then the Global Response Module is activated and the local eSOM map of node D is distributed using the Global Response – Map Distribution Protocol in order to notify all neighbors in the communication range of node D about the attack so that all neighbors remove the attacked node from their routing tables and update the paths they use for their message transmissions.

V. PERFORMANCE EVALUATION AND ANALYSIS

The simulated network has no preexisting infrastructure and that the employed ad hoc routing protocol is AODV. We implemented the simulator within the ns-2 library. Our simulation modeled a network of 50 hosts placed randomly within an 1800 × 1000 m$^2$ area. Each node has a radio propagation range of 250 m and the channel capacity was 2 Mbps. The nodes in the simulation move according to the 'random way point' model. The minimum and maximum speed is set to 0 and 10 m/s, respectively, and pause times at 0, 20, 50, 70 and 200 sec. A pause time of 0 sec corresponds to the continuous motion of the node and a pause time of 200 sec corresponds to the time that the node is stationary.

We evaluated the performance of our proposed intrusion detection engine for 5, 10, 15 and 20 malicious nodes. The malicious behavior is carried between 50 and 200 sec. The nodes perform normally between 0 and 50 sec. On average, twenty traffic generators were developed to simulate Transmission Control Protocol (TCP) data rate to ten destination nodes. The sending packets have random sizes and exponential inter-arrival times. The mean size of the data payload was 512 bytes. Each run is executed for 200 sec of simulation time with a feature-sampling interval of one sec.

We simulated a constant selective packet-dropping attack where the attacker simply discards all data packets while it functions legitimately concerning routing and MAC layer packets. This type of attack is extremely difficult to detect if we consider that packet dropping is due to a malicious behavior or mobility. The malicious node exhibit malicious behavior when it is most advantageous to him and not from the beginning of the traffic.

The statistical features we have used have been introduced by Liu et al. [3] and are the following: *Network allocation vector (NAV), Transmission traffic rate, Reception traffic rate, Retransmission rates of RTS packets, Retransmission rates of DATA packets, Active neighbor node count, Forwarding node count*. We have normalized the data with mean zero and variance one.

For the evaluation we have used the Databionics eSOM tool [11]. The evaluation proves that we can achieve a differentiation between normal and abnormal behaviors concerning packet-dropping attacks. The best matches of the trained dataset and, thus, the corresponding dataset were manually grouped into clusters representing normal and attack behavior. Thus, we identify the regions of the map that represent a cluster that can be used for the classification on new datasets. The eSOM of a trained dataset is depicted in Fig. 2. As it can be clearly seen the training data set has been divided in two classes, normal data class (dark color) and packet dropping data class (light color). To evaluate the efficiency of the proposed intrusion detection engine we use the *Detection rate* and the *False alarm rate*.

Fig. 8 presents the average Detection rate of all source nodes that present traffic activity regarding the used pause times. The detection rate seems not to be influenced by the mobility and in all cases to be over 80%. For long pause times the rate slightly

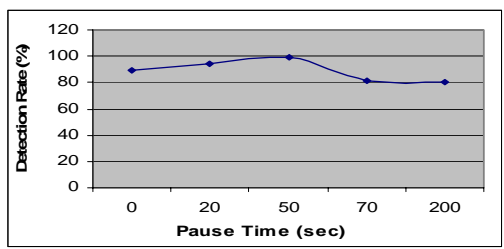

Fig. 8. Detection Rate vs. Pause Time

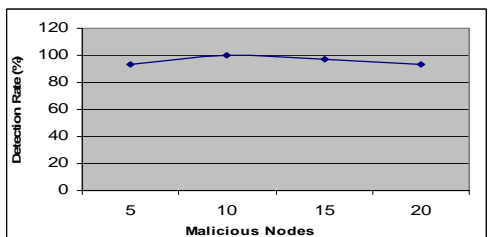

Fig. 9. Detection Rate vs. Number of Malicious Nodes

reduces which is due to the TCP traffic and the degradation of the mobility of the network.

The detection rate as a function of the number of malicious nodes is presented in Fig. 9. The rate is rather high and always over 80%. When few malicious nodes exist in the network, the connections that are influenced by them are also a few. Thus, the detection rate is decreased. When the number of malicious nodes is high compared to the number of source nodes, the TCP connections generated are a few, leading to a decrease in the detection rate.

Tables II and III present the average false alarm rate as a function of the paused times used and the number of malicious nodes, respectively. When a source node generates traffic to different destinations and one of these connections is influenced by malicious nodes, then eSOM finds it difficult to distinguish among normal and abnormal traffic.

Through the proposed GKA protocol, the following security goals are achieved: k*ey secrecy*, *key independence*, *forward secrecy*, *backward secrecy* [8]. *Key Secrecy* is achieved since the key can be computed only by the group members. Moreover, since the disclosure of any set or group keys, does not lead to the disclosure of any other group key, *key independence* is also achieved. The compromise of current session secrets does not imply compromise of future session's secrets. Thus a leaving member is prevented from accessing the group communications and *forward secrecy* is achieved. The disclosure of current session secrets does not imply the disclosure of past session secrets, thus *Backward Secrecy* is also achieved.

TABLE II
FALSE ALARMS VS PAUSE -TIME

| Pause time (sec) | False Alarm (%) |
|---|---|
| 0 | 21 |
| 20 | 20 |
| 50 | 22 |
| 70 | 20 |

TABLE III
FALSE ALARMS VS NUMBER OF MALICIOUS NODES

| Malicious nodes | False Alarm (%) |
|---|---|
| 5 | 26 |
| 10 | 22 |
| 15 | 17 |
| 20 | 21 |

## VI. CONCLUSIONS AND FUTURE WORK

Our intrusion detection engine presents a rather high detection rate and its main advantage is the visual representation of the normal-attack state in a MANET. Moreover, it has the ability to immediately respond in the case of a possible intrusion by selecting the more secure node as indicated by its U-Matrix map for forwarding the information. In order to verify the reliability and avoid possible alteration of the maps the proposed key agreement protocol must be used.